\documentclass[pra,showpacs,floatfix,amsmath,twocolumn,amsfonts]{revtex4} 
\usepackage{graphics}
\usepackage{epsfig}
\usepackage{color}

\begin{document}

\title{Scattering of gap solitons by PT-symmetric defects}

\author{F. Kh. Abdullaev$^1$, V.A. Brazhnyi$^2$, M. Salerno$^3$}
\affiliation{$^1$ Physical - Technical Institute, Uzbek Academy of
Sciences, 2-b, G. Mavlyanov str., 100084, Tashkent, Uzbekistan \\
$^2$ Centro de F\'{\i}sica do Porto, Faculdade de Ci\^encias, Universidade do Porto, R. Campo Alegre 687, Porto 4169-007, Portugal\\
$^3$  Dipartimento di Fisica "E. R. Caianiello",
Universit\'{a} di Salerno, via Giovanni Paolo II, stecca 9,  I-84084, Fisciano (SA), Italy}

\date{\today}
\begin{abstract}
The resonant scattering of gap solitons (GS) of the periodic nonlinear Schr\"odinger equation  with a localized defect which is symmetric under the parity and the time-reversal (PT) symmetry,  is investigated. It is shown that for suitable amplitudes ratios of the real and imaginary parts of the defect  potential the resonant transmission of the GS through the defect becomes possible. The resonances occur for potential parameters which allow the existence  of localized defect modes with the same energy and norm of the incoming GS. Scattering properties  of GSs of different band-gaps with effective masses of opposite sign are investigated. The possibility of  unidirectional transmission and blockage of GSs by PT defect, as well as, amplification and destruction induced by  multiple reflections from two PT defects, are also discussed.
\end{abstract}
\maketitle

\section{Introduction}

Recently it has been shown  that non-Hermitian Hamiltonians that are symmetric with respect to both parity and time-reversal (PT) symmetry can have a fully real spectrum,  in spite of the non-Hermiticity of the Hamiltonian~\cite{Bender}. This observation has attracted the attention of many researchers, both for theoretical developments of dissipative systems in quantum mechanics, and for developments of concrete applications in the fields of  optics~\cite{El-Ganainy}, 
plasmonics~\cite{Ben},  electronics~\cite{Kottos} and  meta-materials~\cite{Lazarides}.

In particular, in the field of nonlinear optics, PT-symmetric potentials  are presently investigated for management of light propagation in media with specific spatial distributions of gain and losses~\cite{El-Ganainy}. In this context, many interesting phenomena  have been reported, including double refraction of beams~\cite{Chr1}, non-reciprocal propagation in periodic PT-symmetric media~\cite{Lin}, existence of optical solitons~\cite{Mussl,AKKZ}, routing in optical PT-symmetric mesh lattices~\cite{Chr3}, etc. PT-symmetric lattices have also been suggested for realizations  in resonant media with three-level atoms~\cite{Kon}.

The scattering of usual continuous and discrete solitons by localized PT potentials have been recently investigated in~\cite{Nazari} for the case of  Scarf II type PT potential,  and in~\cite{Dmitriev} for PT defects in quasi-linear regime where it has been shown that reflected and transmitted small amplitude waves  can be amplified in the scattering process. The possibility of soliton switching in a PT-symmetric coupler induced by the gain and loss properties of the PT defect was also  suggested in~\cite{AKMS}.

Existence and stability of defect-gap solitons in real periodic optical lattices (OL) with PT-symmetric nonlinear potentials have been demonstrated in~\cite{Wang2}. In this context, particular attention has been  devoted to the scattering properties of linear waves propagating in PT-symmetric optical media, as well as to the existence of localized states, both in linear and nonlinear cases.
The existence and stability of gap solitons (GSs) in PT-symmetric lattices with single-sided defects was considered in~\cite{Zhou,Wang} for the continuous case, and for the discrete case with a nonlocal nonlinearity in~\cite{Hu} where it was shown that nonlocality can enlarge soliton existence regions in parameter space.

Scattering of GSs by localized defects has been extensively investigated in the conservative case. In particular,  the existence of repeated reflection, transmission and trapping regions  for increasing  defect amplitudes has been  demonstrated in~\cite{BS2011} where the phenomenon of resonant transmission was discussed and ascribed to the existence of defect modes matching the energy  and the norm of the incoming GS. Moreover, it was shown that the number of  resonances observed in the scattering coincides with  the number of bound states existing inside the defect potential  and that the sign of the effective mass of the GS plays  important role in the interaction with the defect potential~\cite{BS2011}. Scattering properties of GSs by PT-symmetric defect potentials, to the best of our knowledge, have not been investigated. Quite recently, the existence of defect modes of PT-symmetric OL has been  experimentally reported in ~\cite{ExpDefect}.

Possible extensions of the above conservative results to the case of PT defects can be of interest in several respects. In particular, it is interesting to see if the interpretation of the scattering  properties in terms of  resonances with PT defect  modes is still valid.  In addition, the interplay between effective mass, potential amplitudes and interaction is also very interesting to explore in the presence of PT-symmetric defects.

The aim of the present paper is to investigate the scattering properties of a GS of the periodic nonlinear Schr\"odinger equation (NLSE) by localized PT-symmetric defects. In particular, we show that  resonant transmissions of GSs through a PT defect become possible for amplitudes ratios of real and imaginary parts of the PT potential which allow  the existence of defect modes with the same  energy and norm of the incoming GS. For PT defects with a small imaginary part, the scattering properties are found to be very similar to those reported for the conservative case~\cite{BS2011}. As the imaginary part of the PT defect potential is increased, however, we show that  GS can be strongly amplified or depleted especially when potential parameters are very close to higher resonance. Resonant transmission peaks obtained from direct numerical integrations of the NLS equation  are found to be in all cases in good agreement with those predicted by a stationary PT defect mode analysis.

Scattering properties of  GS  with different effective masses are also investigated. In particular, we show that GS with opposite effective mass behave similarly when the sign of the PT defect is reversed, this confirming the validity of an effective mass description in the scattering by PT defects. The possibility of unidirectional transmission of GS through PT defects, and the amplification or destruction of a  GS trapped between two PT defects, are also considered at the end.
Finally, we remark that PT-symmetric potentials are presently  experimentally implemented in optical systems and we expect that the above results can  have experimental implementations in systems similar to the one considered in ~\cite{ExpDefect}.

The paper is organized as follows. In section II we introduce the model equation and discuss the main properties of the system. In section III we present scattering results obtained from direct numerical PDE integrations of the system, for resonant transmissions, trapping and reflections  of a GS through a PT defect, as a function of the potential  parameters.
This is done both for a GS of the semi-infinite gap and for GS of the first band-gap, with positive and negative effective masses, respectively,  and results are compared  with those obtained from defect mode analysis. In Sec. IV  we discuss  possible applications of the scattering properties of a GS both by a single PT defect and by a couple of defects, while in the last section  the main results of the paper are briefly summarized.

\section{The model}
The model equation we consider is the following normalized one-dimensional NLSE
\begin{eqnarray} \label{gpe}
i\Psi_t=-\Psi_{xx}+ \left(V_{ol}(x) + V_d(x)\right)\Psi+ \sigma|\Psi|^2\Psi,
\end{eqnarray}
with $V_{ol}(x)$ denoting a periodic potential (optical lattice) of period  $L$:  $V_{ol}(x)=V_{ol}(x+L)$ and
$V_d$ a localized PT-symmetric complex defect introducing gain and loss in the system.
\begin{figure}[h]
\epsfig{file=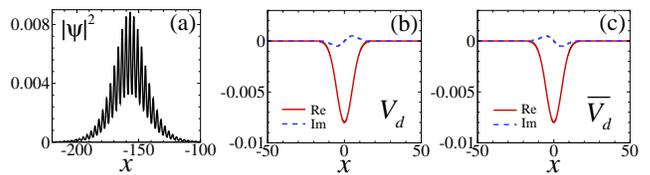,width=8.5cm}
\caption{(Color online) Initial profile of a GS located in the semi-infinite gap at $E_s=-0.125$ [panel (a)], and defect potential $V_d(x)$ with $\xi=0.02|\eta|$ [panel (b)] and with $\xi=-0.02|\eta|$ [panel (c)].  Other parameters are $\eta=-0.1$, $V_0=-1$.}	
\label{ini}
\end{figure}	

This equation arises in connection with  the propagation of a plane light beam in a Kerr nonlinear  media with a linear complex refraction index $n(x) =n_R(x) + i n_I(x)$ introducing periodic modulation and localized gain-loss distribution in the transverse $x$ direction.
As is well known, the wave equation for the propagation of the electric field of the beam,  in the paraxial approximation can be written as
\begin{equation}
\label{NLS}
iE_z + \frac{1}{2\beta}E_{xx} + k_0\left[n_R(x) + i n_I(x) + \sigma|n_2||E|^2\right] E =0,
\end{equation}
where $E(x,z)$ is the electric field, $z$ is the longitudinal (propagation) distance, $\beta = n_0 k_0 = 2\pi n_0/\lambda_0$  the propagation constant, with $n_0$ and $n_2$ the  background and quadratic parts of the refraction index, respectively, and with $\sigma$ fixing the sign of the coefficient of the Kerr nonlinearity (e.g. $\sigma=1$ for focusing and $\sigma=-1$ for defocusing cases).
It is known that in order to satisfy the PT symmetry $n_R(x)$ must be an even function  while the gain-loss component, $n_I(x)$, must be odd. Eq.~(\ref{gpe}) then follows from Eq.~(\ref{NLS}) after introducing dimensionless variables
$t = \frac{z}{L_b}, x = \frac{x}{x_b}$
and the rescaling of the field amplitude and refraction index according to
\begin{equation}
\sqrt{k_0|n_2|L_b}\ E =\Psi,\;\;  2\beta^2 x_b^2 n(x) = V_{ol} + V_d
\end{equation}
(here $x_b$ denotes the initial width of the beam and $L_b = \beta x_b^2$ its diffraction length).
\begin{figure}[h]
\epsfig{file=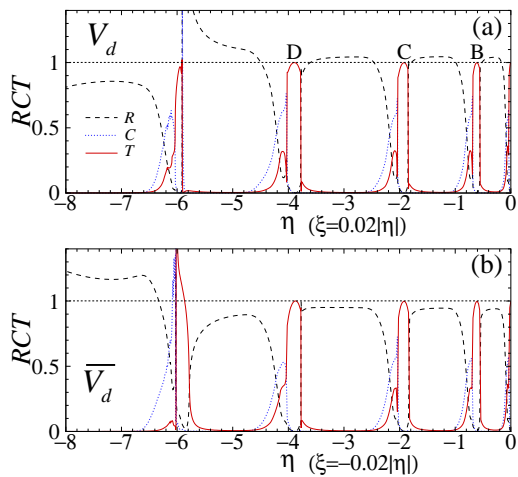,width=8cm}
\caption{(Color online) $RCT$ diagram for $V_d(x)$ with $\xi=0.02|\eta|$ [panel (a)]  and $\overline{V_d}(x)$  [panel (b)] with $\xi=-0.02|\eta|$.  Other parameters: $v=0.05$, $E_s=-0.125$, $V_0=-1$.}	 \label{fig_eta+-}
\end{figure}	
In the following  we fix $V_{ol}=V_0 \cos(2x)$  and take the defect potential $V_d(x)$ of the form
\begin{eqnarray}
\label{PTd}
V_d(x)=  \frac{\eta + i \xi x}{\sqrt{2\pi}\Delta}\exp\left[-(x-x_0)^2/(2\Delta^2)\right], \label{defect}
\end{eqnarray}
where $\eta$ is the strength of the conservative part of the defect while 
coefficient $\xi$ stands for the gain-dissipation parameter. The width of the defect is fixed to $\Delta=5$ in all numerical calculations. Similar PT defect was considered recently also in Ref.~\cite{Wang}. Although in this paper we  mainly concentrate on the case of a single PT defect, some result about the scattering of GSs from two PT defects will also be discussed at the end.

As it is well known, in the absence of any defect potential, Eq.~(\ref{gpe})  possesses  families of exact GS solutions with energy (propagation constant) located in the band-gaps of the linear eigenvalue  problem
\begin{equation}
\label{eigenval}
\frac{d^2 \varphi_{\alpha k}}{d x^2} + \left[ E_{\alpha}(k) - V_{ol}(x)\right]\varphi_{\alpha k}=0,
\end{equation}
where $\varphi_{\alpha k}(x)$ are orthonormal set of Bloch functions with $\alpha$ denoting the band index and $k$ the crystal-momentum  inside the first Brillouin zone (BZ): $k\in [-1,1]$. It is also known  that small-amplitude GSs with chemical potentials $E_s$ very close to band edges are of the form $\psi(x,t)=A(\zeta,\tau)\varphi_{\alpha k}(x)e^{-i E_{\alpha}(k)t}$ with the envelope function $A(\zeta,\tau)$ obeying the following NLSE
\begin{eqnarray}
i\frac{\partial A}{\partial \tau} = -\frac{1}{2M_{eff}} \frac{\partial^2 A }{\partial \zeta^2} + \chi |A|^2A
\label{eq:env}
\end{eqnarray}
where  $\tau$ and $\zeta$ are slow temporal and spatial variables,  $M_{eff}=(d^2E_\alpha/dk^2)^{-1}$ denoting the soliton effective mass and $\chi=\sigma\int |\varphi_{\alpha k}|^4 dx $ the effective nonlinearity~\cite{KS}.
The condition for the existence of such solitons is  $\chi M_{eff} <0$ \cite{SKB08} and coincides with the condition for the modulational instability of Bloch wavefunctions at the edges of the BZ ~\cite{KS}. In the presence of an OL with a localized PT defect, the linear spectral problem will still display a band structure but with additional localized states (defect modes) that are associated to real eigenvalues (in band gaps) when the imaginary part of potential is below a critical value  $|\xi_c| = |\eta|/\sqrt{2}\Delta$. Above this point, defect mode  spectrum becomes mixed  with complex pairs of eigenvalues, this corresponding to a  dynamical breaking of the PT-symmetry~\cite{Ahmed}. We remark that in nonlinear optics, PT-symmetry and PT-symmetry breaking have been both observed experimentally~\cite{Ruter,Guo}.
\begin{figure}[h]
\epsfig{file=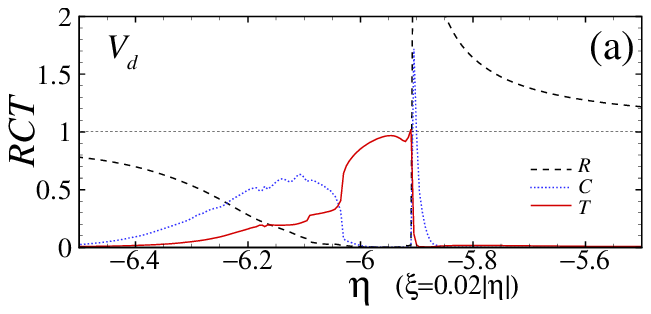,width=8cm}
\epsfig{file=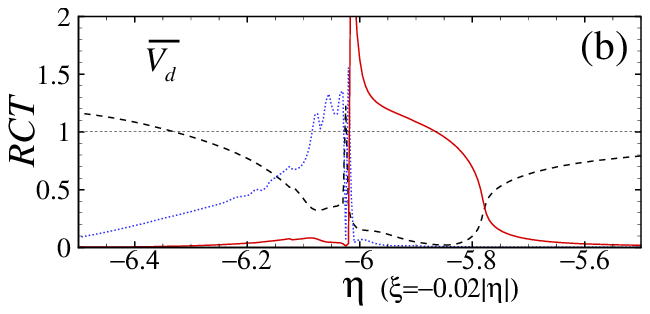,width=8cm}
\caption{(Color online) Zoom of Figs.~\ref{fig_eta+-}(a),(b) showing details in the interval $\eta\in[-6.5, -5.5]$ around the resonance.}	
\label{fig_eta+-zoom}
\end{figure}

\section{Scattering of GS by a PT defect: numerical results}

In order to investigate scattering properties of a GS by a localized PT defect, we compute by means of direct numerical integrations of Eq.~(\ref{gpe}) the transmission ($T$), trapping ($C$) and reflection ($R$) coefficients  defined as
\begin{eqnarray}
&& T=\frac{1}{N_0}\int_{x_c}^{\infty}|\Psi(x,t_s)|^2 dx, \nonumber \\
&& C=\frac{1}{N_0}\int_{-x_c}^{x_c} |\Psi(x,t_s)|^2 dx, \\
&& R=\frac{1}{N_0} \int_{- \infty}^{-x_c} |\Psi(x,t_s)|^2 dx, \nonumber
\end{eqnarray}
where the integrals are evaluated after a sufficient long time, $t_s$  (typically $t_s \approx 20000$), for the process to become stationary.
Here $N_0$ denotes the initial norm of the incoming GS (e.g. $\int_{- \infty}^{\infty} |\Psi(x,t=0)|^2 dx$), and the interval $[-x_c, x_c]$ represents the trapping region around the PT defect,  with $x_c$ fixed in all our calculations to $x_c= 30 L$. In particular, we are interested to characterize the dependence of the above  coefficients on the PT defect parameters, $\eta$ and $\xi$, both for a GS of the semi-infinite gap and for a GS of the first band-gap, having positive and negative effective masses, respectively.
\begin{figure}[h]
\epsfig{file=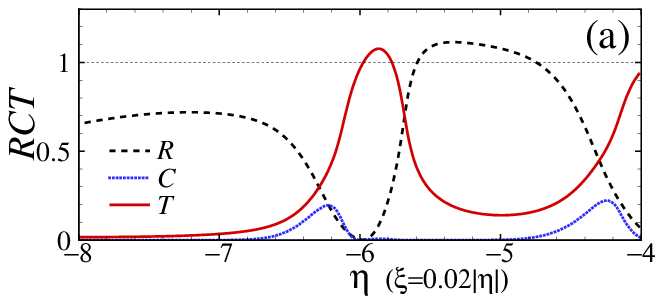,width=8cm}
\epsfig{file=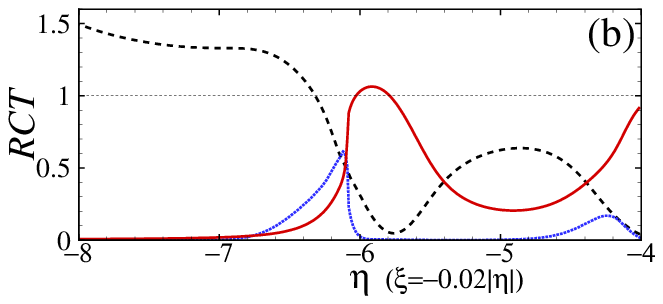,width=8cm}
\caption{(Color online) The same as in Fig.\ref{fig_eta+-zoom} but for an incoming GS velocity  $v=0.1$. Other parameters are fixed  as in Fig.~\ref{fig_eta+-}(a),(b). }	
\label{fig_eta+-zoom_v01}
\end{figure}
Notice that differently from the conservative case, the sum of the above coefficients is not normalized to 1 , e.g. $R+T+C \neq 1$, due to the presence of gain and loss in the system which does not allow the norm conservation. In particular, the above coefficients   during the scattering can become larger  than one due to the gain action of the PT defect. In all numerical investigations reported below, the GS is constructed as a stationary solution of the periodic NLS equation located at large distance ($\approx 100 L$)  from the PT defect (far away from the defect such states practically coincide with those of the NLSE with a perfect OL). The stationary GS is then put in motion  by means of the phase imprinting technique, e.g. by applying  a linear phase $e^{-i\sigma vx/2}$ to the stationary wavefunction.

\subsection{GS of the semi-infinite gap}

Let us first consider the case of a GS  of the semi-infinite gap, e.g. with $\sigma=-1$ in Eq.~(\ref{gpe}),  with energy (propagation constant)  $E_s=-0.125$ close to the bottom edge of the lowest energy band.
Initial GS profile and PT defect shape $V_d(x)$ are shown in Fig.~\ref{ini} for the case  $\xi= \pm 0.02|\eta|$. In the numerical experiment we apply an initial velocity to the GS, typically in the range $0.02 \div 0.1$,  and gradually decreasing the strength of the defect parameter  $\eta$ under condition  $\xi=\pm 0.02|\eta|$,  in order to obtain the $RCT$ coefficients depicted in Figs.~\ref{fig_eta+-}(a),(b). We see that for weak defect amplitudes  and for the same GS initial velocity ($v=0.05$),  the positions of the $T$-peaks, labeled B, C, and D in panel (a), mostly coincide with the ones of the conservative case considered in~\cite{BS2011} (see Fig.3 in~\cite{BS2011}).

It is worth to note the differences in the behavior of the reflection coefficient. While in the case ($\xi=0$) the coefficient $R$ approaches the value 1 in the regions of non existence of defect modes, one can see that  in the case $\xi=0.02|\eta|$ the $R$ coefficient in the interval $\eta\in[-6,0]$  in the total reflection regions becomes slightly greater than 1 [see Fig.~\ref{fig_eta+-}(a)], meaning that during reflection the GS has been  amplified by the defect. The opposite behavior is observed for the case $\xi=-0.02|\eta|$ (corresponding to the defect $\overline{V_d}(x)$), e.g. in the  reflection regions inside the interval  $\eta\in[-6,0]$ the reflection coefficient  is always smaller than 1, meaning that the GS has been damped during the reflection [see Fig.~\ref{fig_eta+-}(b)].
\begin{figure}[h]
\epsfig{file=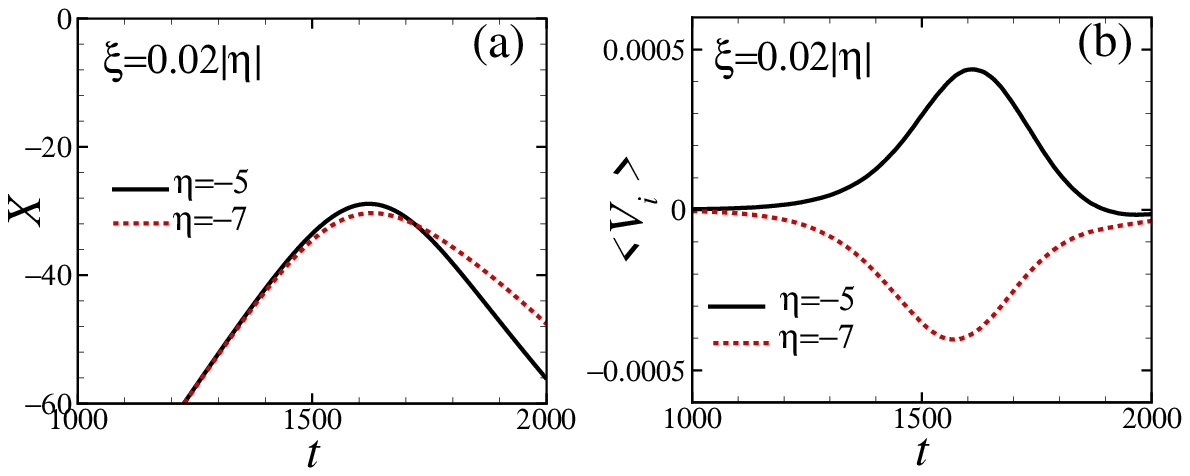,width=8cm}
\epsfig{file=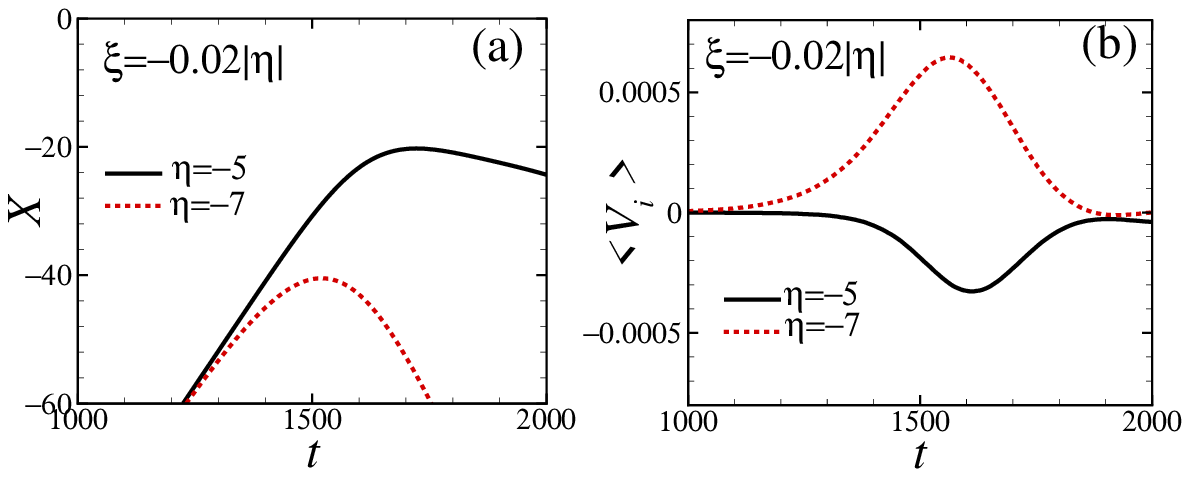,width=8cm}
\caption{(Color online) Trajectories of the center of the density distribution $X(t)$ (left panels) and mean imaginary part of PT defect $\langle V_i \rangle $ (right panels) of a GS during reflection. Top row corresponds to case $\xi=0.02|\eta|$ and bottom  row to  case $\xi=-0.02|\eta|$. Incoming GS velocity and other parameters are fixed as in Fig.~\ref{fig_eta+-zoom_v01}.
}	
\label{fig_conv}
\end{figure}

This behavior of the $R$ coefficient may at a first sight appear counter-intuitive, especially if one observes  that in our numerical experiments the GS  is always coming from the left and when it gets  amplified (resp. depleted) it arrives first at the loss (resp. gain) side of  the defect, from which  one could expect just the opposite, e.g. a depletion (resp. amplification) of the GS from the defect. The observed behavior, however, can be understood if one consider in more detail  the GS dynamics during reflection.
From an intuitive point of view one can argue that since for $\xi>0$ (resp. $\xi<0$) the GS  interacts first  with the loss (resp. gain),  it can passes this region with some velocity  so that  the  turning point of its  dynamics occurs more close to the  gain (resp. loss) region of the defect (this is particularly true if  the initial velocity is high or the imaginary part of the defect is small).
Considering that the GS is an extended object and  for the cases considered in this paper its typical width is of about $30 L$ (see Fig.~\ref{ini})  e.g. much larger than the size of the defect with a width $\approx 8 L$, this means that  during the reflection the GS will be in any case exposed  to the action of the  gain  (resp. loss) side of the defect and the influence of this region on the  dynamics will be larger as closest will be the turning point at the origin. To understand if the GS will emerge 
amplified or  depleted  from the reflection it is convenient to consider  the mean imaginary part of the defect potential seen by the GS at a given time defined as
\begin{equation}
 \langle V_i \rangle (t) = \frac 1 {N_0} \int_{-\infty}^{\infty} Im[V_d(x)] |\Psi(x,t)|^2 dx.
\label{conv}
\end{equation}
It is clear that if  $\int \langle V_i \rangle dt$ is positive (resp. negative), the amplification (resp. depletion) of the GS  is expected during the reflection.
\begin{figure}[h]
\epsfig{file=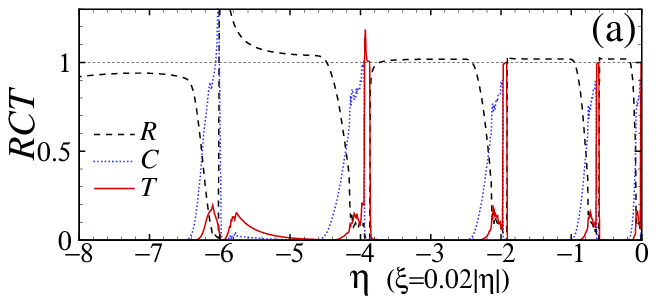,width=9cm}
\epsfig{file=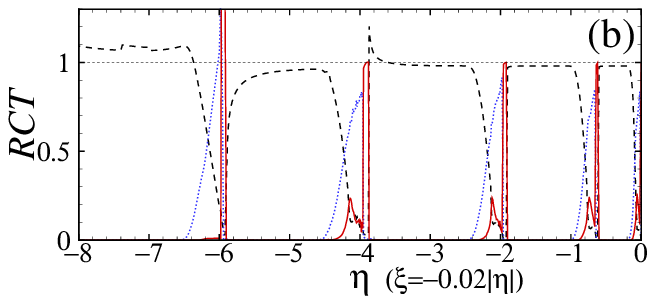,width=9cm}
\caption{(Color online) $RCT$ diagram for $\xi=0.02|\eta|$ (panel a)  and $\xi=-0.02|\eta|$  (panel b).  Other parameters: $v=0.02$, $E_s=-0.125$, $V_0=-1$. }	\label{fig_eta+-v002}
\end{figure}	
\begin{figure}[h]
\epsfig{file=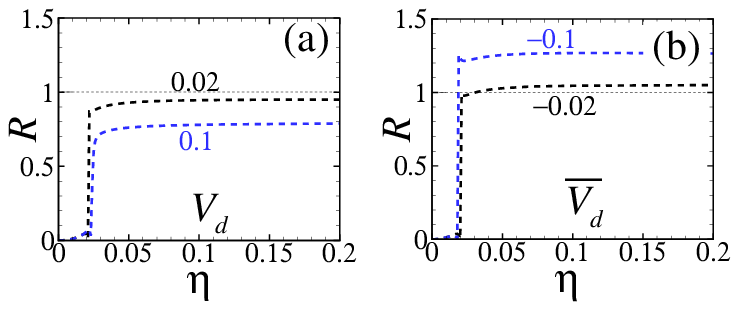,width=8cm}
\caption{(Color online) Behavior of the $R$ coefficient in the scattering of a GS of the semi-infinite band gap  with $v=0.05$, by a PT defect with $\eta>0$. The ratio $\xi/|\eta|$ is indicated near the corresponding curve.}		 \label{fig_xi+-00_eta+}
\end{figure}

This is exactly what it is shown in the right panels of Fig.~\ref{fig_conv} where results of two distinctive cases from Fig.~\ref{fig_eta+-zoom_v01}, with $\eta=-5$ and $\eta=-7$, are reported.
In the left  panels of Fig.~\ref{fig_conv} we have  depicted the trajectory of the center  $X$
\begin{equation}
 X(t)= \frac 1 {N_0} \int x |\Psi(x,t)|^2 dx
\label{c_mass}
\end{equation}
of the density distribution during the reflection.
One can see from this figure that, in agreement with our intuitive argument, for smaller values of $|\eta|$ (e.g. on the right side of resonanceat $\eta \approx -6$)  the GS can  penetrate the defect more and in the cases in which  the  GS is  amplified, the turning point of the trajectory always occurs closer (resp. less close)  to the origin for  $\xi>0$ (resp.  $\xi<0$). The opposite behavior is observed for a GS that is depleted during the reflection.

Interesting results are also observed when the imaginary part of the defect potential is increased and the non-Hermitian character  of the interaction contributes more significantly to the scattering. For the chosen ratio $\xi/|\eta|=0.02$, this occurs around  the value $\eta\leq -6$ as one can see from the details depicted  in Fig.~\ref{fig_eta+-zoom}(a). 
\begin{figure}[h]
\epsfig{file=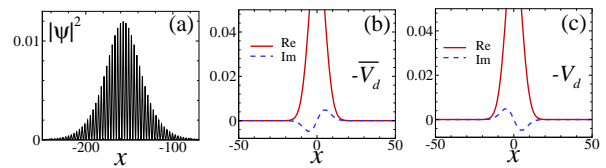,width=8cm}
\caption{(Color online) Initial profile of a GS located near the top of the lower band  at $E_s=0.475$ [panel (a)], and PT defect potential  $V_d(x)$ for $\xi=0.02|\eta|$ [panel (b)] and $\xi=-0.02|\eta|$ [panel (c)].  Other parameters are $\eta=1$, $V_0=-1$.}	 \label{fig0475_ini}
\end{figure}	
From this it is clear that the interaction of the GS with the loss and gain parts of the PT defect changes character when passing through  the resonance point. In particular, one can see that near $\eta=-5.9$ the reflection coefficient shows a rapid growth  corresponding to a strong amplification during the reflection. The explanation of this follows from the same  arguments  given above and from the fact that the interaction with  almost resonant stationary defect modes will further prolong the interaction time that the GS has with the gain side of the defect, this resulting in an higher amplification. This effect can be observed also at lower resonances by decreasing the incoming GS velocity, as one can see from Fig.~\ref{fig_eta+-v002} for the rapid grow of $T$ and $R$ coefficients occurring around $\eta=-4$.

From Figs.~\ref{fig_eta+-}(a),(b), is also quite evident the crossover of the R coefficient from $R>1$ (resp. $R<1$) to $R<1$  (resp. $R>1$) occurring for the case  $\xi=0.02|\eta|$ (resp. $\xi = -0.02 |\eta|$) when $|\eta|$ is increased through  the resonant point $\eta=-6$. This change of behavior can be understood from  the fact that by further increasing  the imaginary part of the PT defect (as is the case when $|\eta|>6$), one reaches the point  in which the turning point of the GS dynamics will always occur in the defect side from where the soliton arrives, so that it is always depleted by $V_d$ and amplified by $\overline V_d$. This explanation also correlates  with the above arguments in terms of turning points and mean effective potentials $\langle V_i \rangle$.

From the more detailed Fig.~\ref{fig_eta+-zoom},  it appears evident that just beyond the point $\eta=-6$,  trapping becomes dominant and due to strong interaction with  defect modes, the GS  becomes very unstable this leading to the irregular oscillatory behavior observed for the trapping coefficient in the  panel (b) of the figure.
By decreasing the velocity of the incoming GS, however,   the transmission peaks become narrow (see Fig.~\ref{fig_eta+-v002})  and the $R$ coefficient becomes closer to 1 in the total reflection regions (scattering is less affected by the complex potential). This is a consequence of the fact that for a smaller velocity a small amount of the GS wavefunction  penetrates the defect and the interaction  with the complex part of the potential is reduced.

It is interesting to discuss also the case $\eta>0$ for which the  real part of the PT defect corresponds to a barrier potential rather than a potential well. 
\begin{figure}[ht]
\epsfig{file=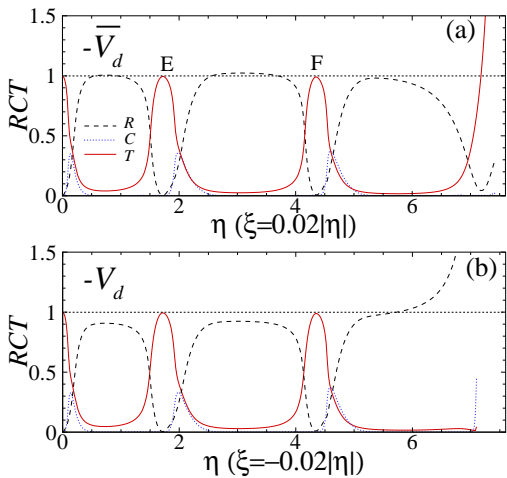,width=8cm}
\caption{(Color online) $RCT$ diagram for  $\xi=0.02 \eta$ [panel (a)]  and $\xi=-0.02\eta$ [panel (b)].  Other parameters: $v=0.05$, $E_s=0.475$, $V_0=-1$.}	\label{fig0475_eta+-}
\end{figure}	
This  obviously does not allow the formation of any stationary  mode inside the defect since in this case  $C=0$ and only transmissions or reflections of the GS are possible. For a conservative defect (e.g. for $\xi=0$) it was shown in \cite{BS2011} that for large defect amplitudes the incoming GS is always totally reflected (e.g. $R=1$ and $T=C=0$). For a PT defect with $\eta>0$, we find that while the  transmission and trapping coefficients continue to be zeros for large $\eta$, the reflection coefficient, in accordance to our  previous discussion, depends on the sign of $\xi$ (as well as  on the ratio $\xi/|\eta|$) and can be smaller or larger than $1$ (see Fig.~\ref{fig_xi+-00_eta+}) depending on whether the GS is interacting more with the dissipative or with the gain side of the defect, respectively.

\subsection{GS of first band-gap}

Scattering properties of a GS belonging to the first band gap in the case of self-focusing Kerr   nonlinearity [$\sigma=1$ in Eq.~(\ref{gpe})]  are quite similar to the ones discussed above. In this  case, however, it is possible to have GS with a negative effective mass if the Kerr nonlinearity is defocusing. To investigate the effects of a negative GS mass on the scattering properties we consider a GS  of energy (propagation constant) $E_s=0.475$  close to the top edge of the lowest band. The initial GS profile and shapes of defect potentials are depicted in Fig.~\ref{fig0475_ini}.
For parameters of the PT potential that are below the threshold of the spontaneous PT-symmetry breaking (as is the case considered here) the spectrum is entirely real with a band structure that is only slightly affected by the defect. Since the effective mass is related to the curvature of the band we expect that an effective mass description of the GS dynamics should still be valid, at least for PT defects quite localized and with imaginary parts not too large. We remark here, however,  that a proof of the validity of the effective mass theorem for periodic  PT potentials is presently lacking (notice that in our case the OL is real and the PT symmetry is only coming from the defect).
\begin{figure}[ht]
\epsfig{file=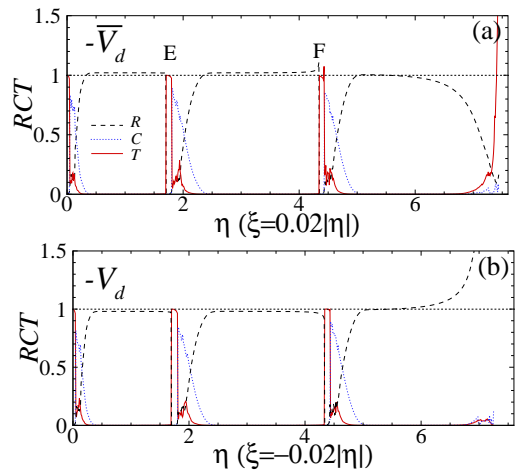,width=8cm}
\caption{(Color online) The same as in Fig.~\ref{fig0475_eta+-} but for a smaller incoming GS velocity $v=0.02$.
 Other parameters are fixed as in Fig. \ref{fig0475_eta+-}.
}	\label{fig0475_eta+-_v002}
\end{figure}	
\begin{figure}[ht]
\epsfig{file=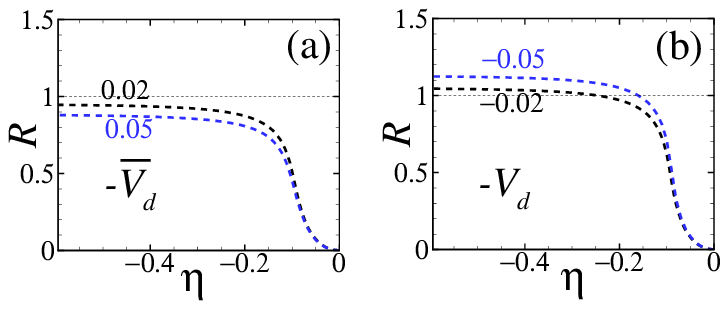,width=8cm}
\caption{(Color online) Behavior of the $R$ coefficient for the scattering of a GS of the first band gap  with negative effective mass, $v=0.05$, by a PT defect with $\eta<0$. The ratio $\xi/|\eta|$ is indicated near the corresponding curve.}	 \label{fig_1g_xi+-00_eta-}
\end{figure}
In an effective mass description one would expect that a change of sign in the effective mass can be compensated by  a change of sign of the defect potential. If true, this would imply that
the scattering properties of a GS with a positive effective mass by a PT defect potential $V_d$ should be similar to those of  a GS with negative effective mass scattered by a defect of opposite sign  $-V_d$.

To check if this is true, we have applied  an initial velocity to GS ($v=0.05$) and constructed as in the previous cases the $RCT$ coefficients as a function of the defect strength. The results are presented in  Fig.~\ref{fig0475_eta+-} for the cases $\xi/|\eta|=0.02$ [see panel (a)] and $\xi/|\eta|=-0.02$ [see panel (b)].
\begin{figure}[h]
\epsfig{file=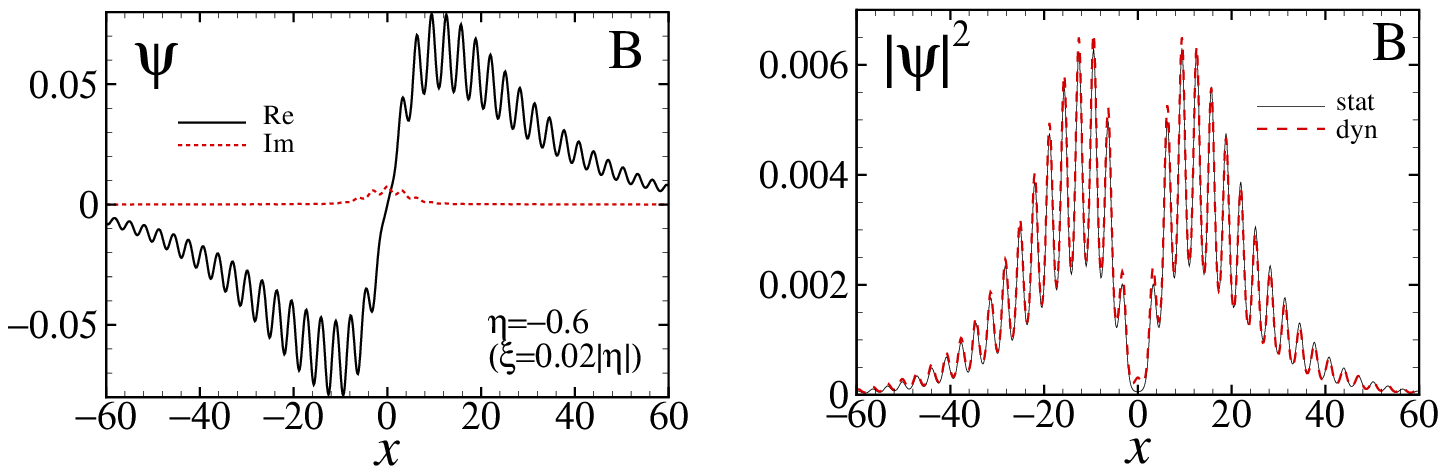,width=8cm}
\epsfig{file=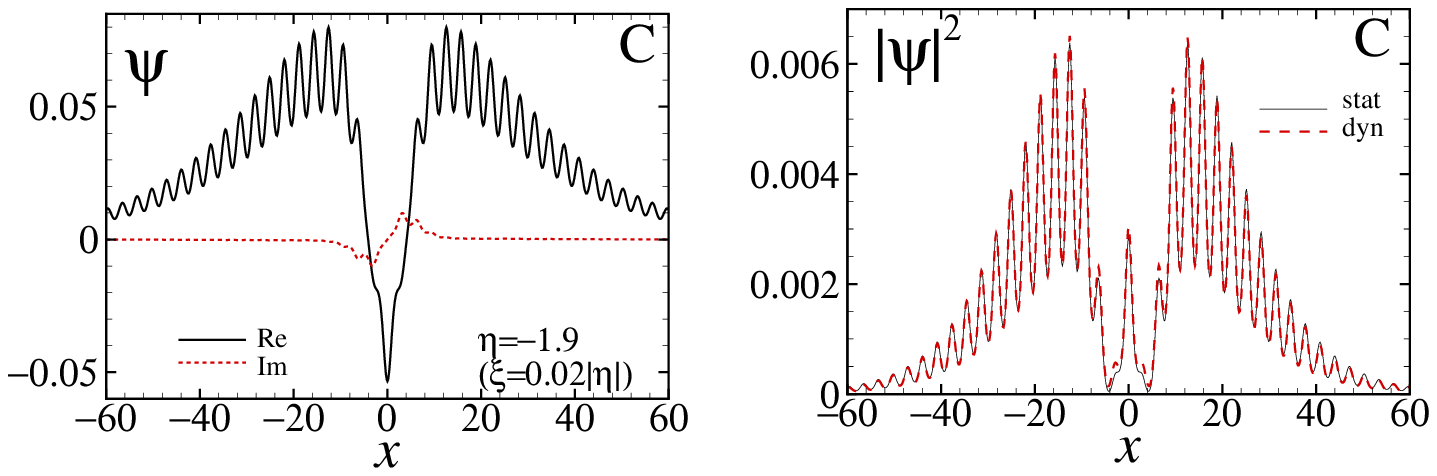,width=8cm}
\epsfig{file=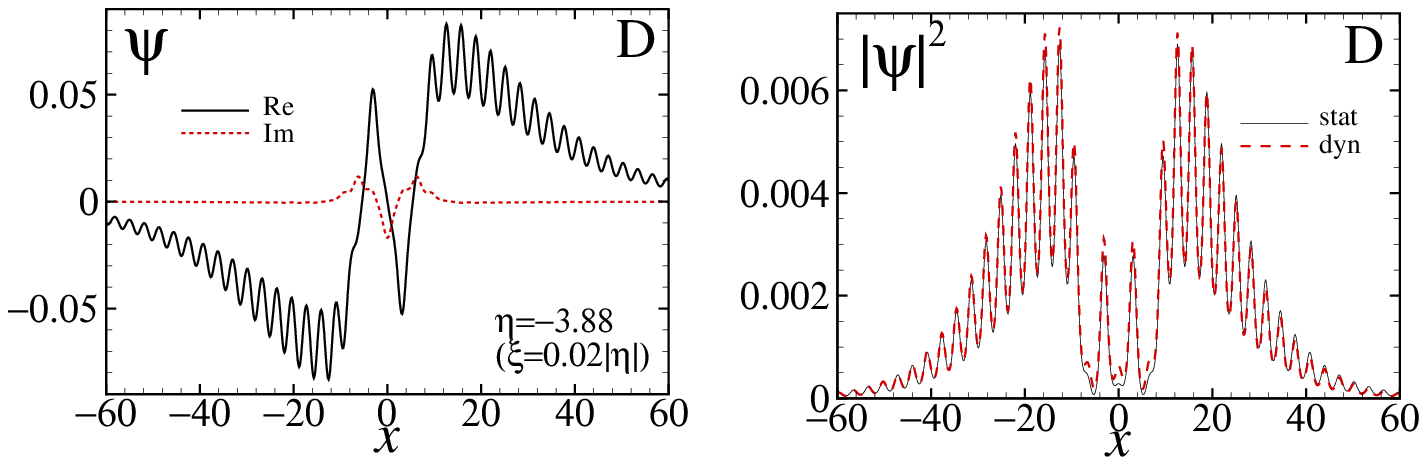,width=8cm}
\epsfig{file=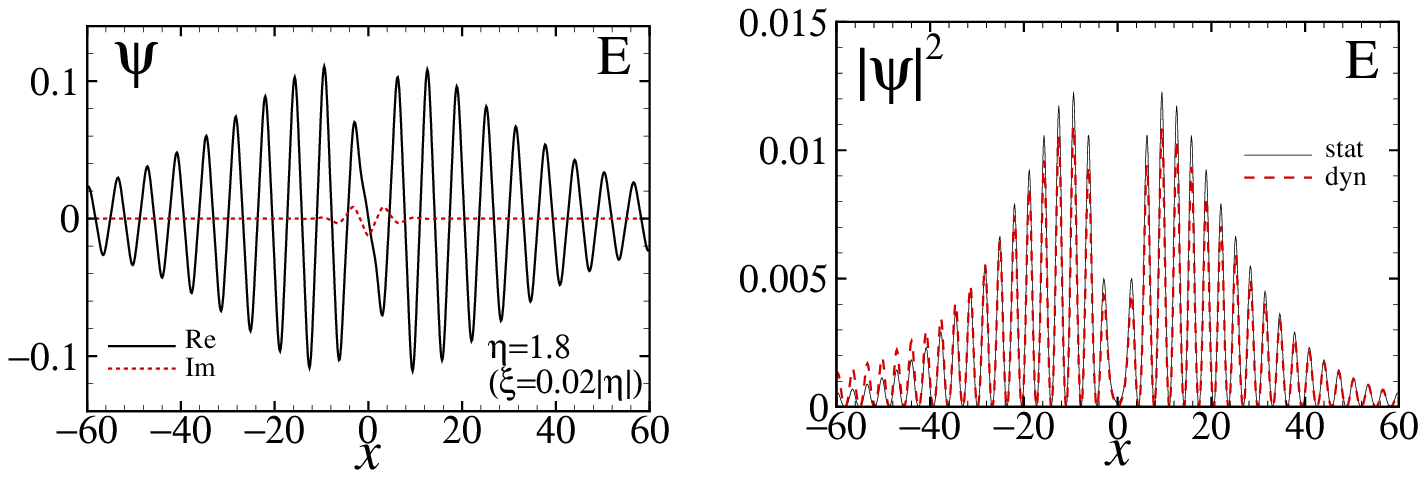,width=8cm}
\epsfig{file=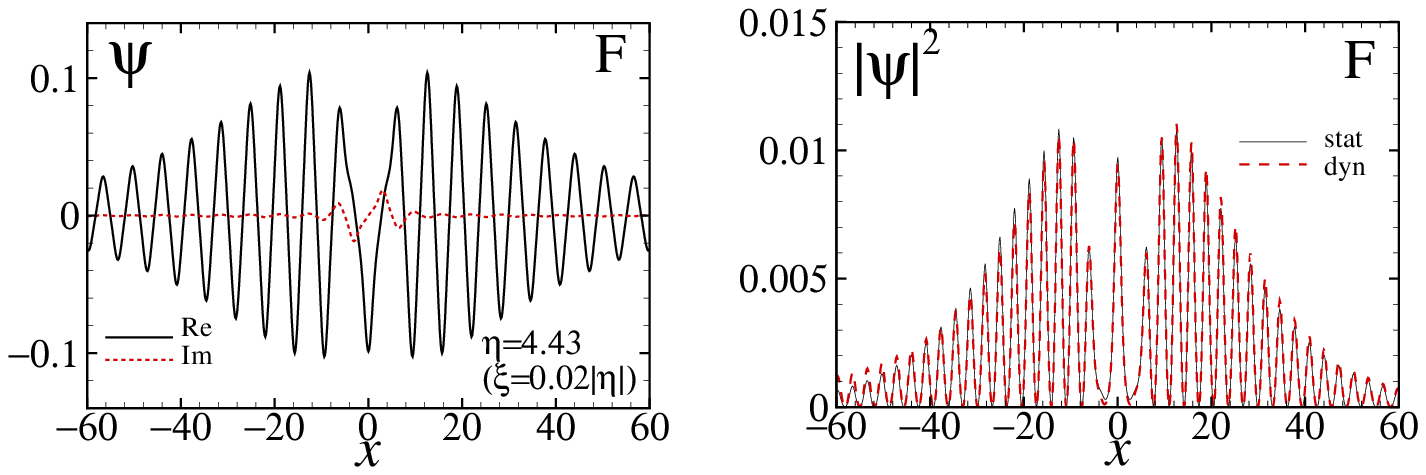,width=8cm}
\caption{(Color online) Top three rows panels. Stationary defect modes at the matching points B, C, D of the Fig.~\ref{fig_eta+-}. Bottom two rows panels.  Stationary defect modes at the matching points E, F of Fig.~\ref{fig0475_eta+-}.
In the left column the real (solid black) and imaginary (dashed red) parts of the defect mode is presented and in the right column its density (solid black) compared to the defect mode calculated numerically (dashed red).
In each row the defect modes at matching points are shown.
}	\label{modes}
\end{figure}

By comparing the $RCT$ diagram of the GS in semi-infinite gap with defect $V_d$ [see Fig.~\ref{fig_eta+-}(a)] with the one in the first gap with defect $-V_d$ [see Fig.~\ref{fig0475_eta+-}(b)], we see that, as expected,  the scattering coefficients behave quite similarly in the two cases, except for the opposite behavior of the  $R$ coefficient at small values of  $|\eta|$ (notice that $R$ is slightly larger than 1 for the GS in  the semi-infinite gap and smaller than 1 for the GS in the first gap). In particular, notice  the rapid  grow of the reflection coefficient $R$ as one approaches the higher resonance in both cases. The discrepancy observed in the behavior of $R$ for small values of $|\eta|$ can be ascribed to the different sizes of the   GS in the two cases, as one can see from the panels (a) of Fig.~\ref{ini} and Fig.~\ref{fig0475_ini}, respectively. The fact that the GS is wider in the first gap permits, for the same incoming velocity, a stronger  interaction with the right  side of the defect, than the one of the more localized GS in the semi-infinite gap. Since the right side of the defect is of loss type for the GS in the first gap (see panel (c) of Fig.~\ref{fig0475_ini}) and of gain type for the GS in the semi-infinite gap (see panel (b) of Fig.~\ref{ini}), this explains the observed discrepancy. Notice that this discrepancy is reduced by reducing the incoming GS velocity [compare Fig.~\ref{fig_eta+-v002}(a) with Fig.~\ref{fig0475_eta+-_v002}(b)]. This can be understood from the reduction at small velocities of the interaction of the GS with the right side of the defect and from the smallness of $\eta$ making the situation close to the one of the conservative case.
Also notice that the decreasing of the incoming GS velocity (see Fig.~\ref{fig0475_eta+-_v002}) leads to the same effects discussed for GS from semi-infinite gap (shrinking the transmission lines and approaching to 1 of the $R$-coefficient in the total reflection regions).

A similar situation is observed for the case $\xi=-0.02|\eta|$ (e.g. for potentials $\overline V_d$ and $-\overline V_d$) with the only difference that the discrepancy at small $|\eta|$ now is of opposite type and the rapid growth at the high resonance occurs for the $T$ coefficient instead than for $R$ as one can see by comparing  Fig.~\ref{fig_eta+-}(b)  with Fig.~\ref{fig0475_eta+-}(a) (also compare Fig.~\ref{fig_eta+-v002}(b) with Fig.~\ref{fig0475_eta+-_v002}(a) for the case of a smaller velocity).

We have also investigated the scattering properties of a negative mass GS by a PT defect with $\eta<0$ (see Fig.~\ref{fig_1g_xi+-00_eta-}). Notice, that due to the negative effective mass, the potential well corresponding to the real part of the PT defect will be seen by the GS as a potential barrier. This case should be then compared with the case $\eta>0$ previously considered for the GS of the semi-infinite gap. Indeed, we find while transmission and trapping coefficients are  zeros the reflection coefficient, in accordance to our  previous discussion for a GS in the semi-infinite gap, depends on the sign of $\xi$ and can be smaller or larger than $1$ as one can see in Fig.~\ref{fig_1g_xi+-00_eta-}. By comparing Fig.~\ref{fig_1g_xi+-00_eta-} with the corresponding Fig.~\ref{fig_xi+-00_eta+},  we see that a part for the discrepancy discussed before and ascribed to the different sizes of the GS, the behavior is in qualitative good  correspondence with what expected from an effective mass description for two GSs of opposite effective masses.

From the above results we conclude that GSs with opposite effective masses, behave quite similarly in the presence of PT defect potentials of opposite signs, this being especially true  for parameters values close to the high resonances.

\subsection{Resonant transmission and PT defect mode analysis}

To check the relevance  of defect modes in the resonant transmission of a GS through  a PT defect, we have explicitly calculated defect modes by solving  the stationary eigenvalue problem associated to Eq.~(\ref{gpe}), and then compared results with those obtained by direct numerical integrations. This is reported in Fig.~\ref{modes} from which we see that there is a good agreement between stationary defect mode analysis and dynamical calculations.

\begin{figure}[ht]
\epsfig{file=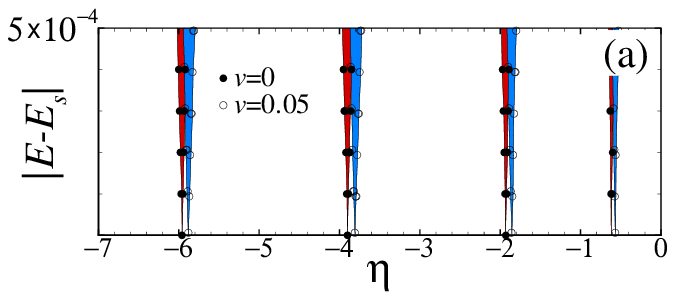,width=8cm}
\epsfig{file=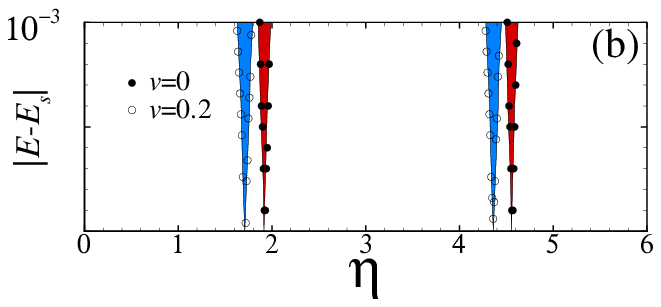,width=8cm}
\caption{(Color online)
Energy mismatch $|E-E_s|$ between defect mode and incoming GS energies  {\it versus} $\eta$,
for $\xi/|\eta|=0.02$, $E_s=-0.125$ (top panel) and  $E_s=0.475$ (bottom panel). Incoming velocities are  $v=0$ (red, filled circles), $v = 0.02$ (blue, open circles) for the top panel,  and $v=0$ (red, filled circles), $v = 0.2$ (blue, open circles) for the bottom panel.
 Other parameters are fixed as in Figs.~\ref{fig_eta+-v002} and \ref{fig0475_eta+-_v002}.
}	 \label{E-Es_sig-1gap}
\end{figure}

\begin{figure}[ht]
\epsfig{file=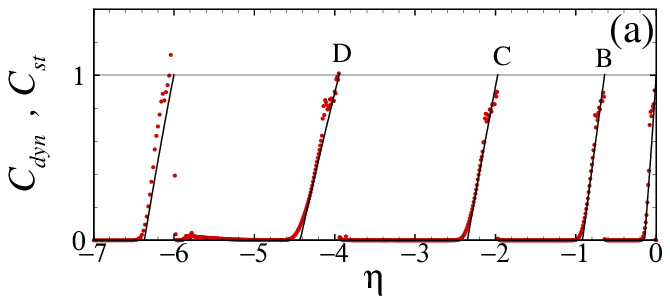,width=8cm}
\epsfig{file=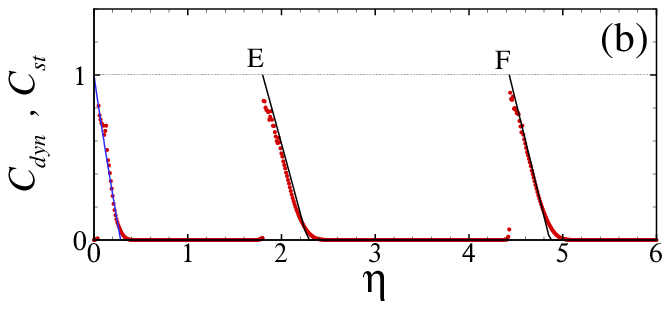,width=8cm}
\caption{(Color online) Trapping coefficient {\it vs} $\eta$ corresponding to
cases considered in corresponding panels of Fig.~\ref{E-Es_sig-1gap}
The dynamical coefficient $C_{dyn}$ (dotted red) refers to the case shown in Figs.~\ref{fig_eta+-v002}  and \ref{fig0475_eta+-_v002} for  $v=0.02$ while  $C_{st}=N/N_0$ corresponds to the norm of defect modes normalized to the initial norm $N_0$ of the incoming GS. 
}	
 \label{CC_sig-1gap}
\end{figure}

Second, we have checked that in all the considered cases the positions of the peaks observed in the $RCT$ diagrams, occur in correspondence with potential parameters that allow the existence of defect modes  with the same energy and norm of the incoming GS (see Figs.~\ref{E-Es_sig-1gap} and~\ref{CC_sig-1gap}).
In particular, in Fig.~\ref{E-Es_sig-1gap} we show the energy mismatch at the resonances  between GS and defect mode for two different cases, while in Fig.~\ref{CC_sig-1gap} we show, for corresponding cases,  the behavior of the stationary and dynamical trapping coefficients as a function of $\eta$. We see from these figures that the agreement between mode analysis and numerical calculations is quite good both for the  energy mismatch and for norms. In particular, notice that positions of peaks is in good agreement even for higher resonances where the imaginary part of the PT defect is not small, this confirming the validity  of the defect modes interpretation for the resonant  transmission of a GS through  PT defects.
\begin{figure}[h]
\epsfig{file=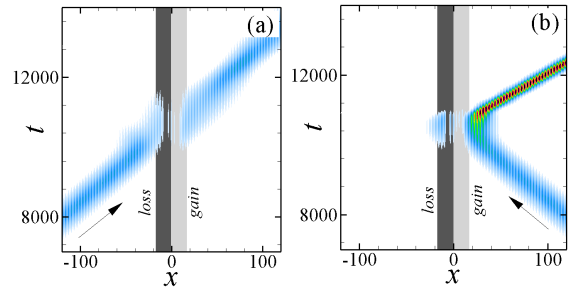,width=8cm}
\caption{(Color online) Contour plots of the GS dynamics. Defect and GS parameters: $\eta=-5.86$, $\xi/|\eta|=0.02$, $E_s=-0.125$, $|v|=0.05$.
}	\label{diode}
\end{figure}
\section{Multiple GS scattering by two PT defects}

In this section we explore the non reciprocity (spatial asymmetry) of the resonant transmission~\cite{diode,nod} that could be used for an unidirectional transmission/blockage of a GS through a PT defect ({\it diode effect}). For this, we fix parameters of the defect potential in the region where it is possible to have total reflection (transmission) for the specific ratio value  $\xi/|\eta|=0.02$ ($-0.02$). Also we refer to the specific case of a GS located in the  semi-infinite gap depicted in Fig.~\ref{diode} (similar results can be obtained for GS of higher band-gaps). For the above fixed ratio it is possible to observe asymmetric (nonreciprocal) behavior at $\eta=-5.8$ (see Fig.~\ref{fig_eta+-zoom}). The results of the interaction of the GS  coming from the left [panel (a)] and from the right [panel (b)] with the PT defect are shown in Fig.~\ref{diode}. As one can see from panel (a), the total transmission of the GS occurs when the GS is coming from the left, while the total reflection with amplification and acceleration is observed when GS comes from the right.

\begin{figure}[h]
\epsfig{file=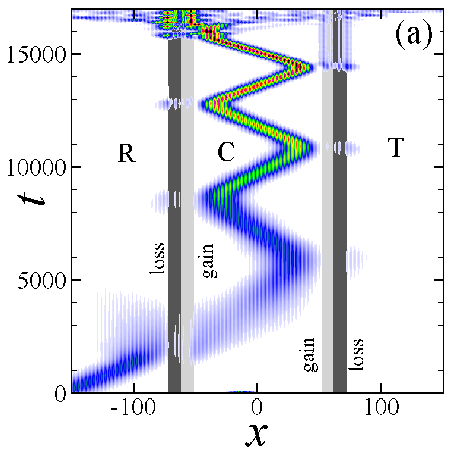,height=5cm}\epsfig{file=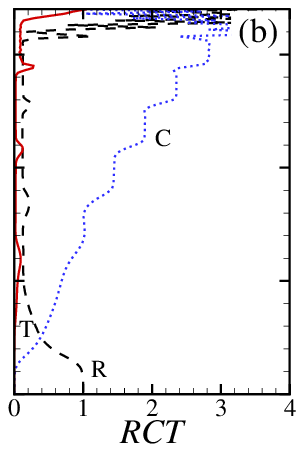,height=5.1cm}
\caption{(Color online) Contour plots of the GS dynamics (left panels) and time evolution  of RTC-coefficients (right panels) for a  GS trapped between two adjacent PT defects with opposite facing gain sides and for parameter values: $x_{1,2}=\pm 20\pi$, $\eta=-5.8$, $\xi_{1,2}/|\eta|=\pm0.02$. Parameters of the initial GS are: $E_s=-0.125$, $v=0.05$.
Notice the amplification of the GS at each reflection and the instability with emission of radiation which appear at late stages.
}	
\label{diode_2}
\end{figure}

By placing  two PT defects symmetrically at $x_{1,2}=\pm 20\pi$ with opposite sign of imaginary part $\xi_{1,2}=\pm 0.02|\eta|$ we obtain that a launched GS from the left enters the intra defects region  and starts to be reflected from  both defects with amplification. The density plot of the dynamics is shown in Fig.~\ref{diode_2}(a) and the dynamics if the $RCT$ coefficients is shown in the panel (b).

As one can see from Fig.~\ref{diode_2}(b) the dynamics of the $C$ coefficient has step like behavior  at each reflection in the region between the two defects the GS being amplified and becoming more localized, this eventually leading to the instability of the GS with emission of waves.
This configuration of PT defects can be seen as a kind of  parametric amplifier for the GS.
\begin{figure}[h]
\epsfig{file=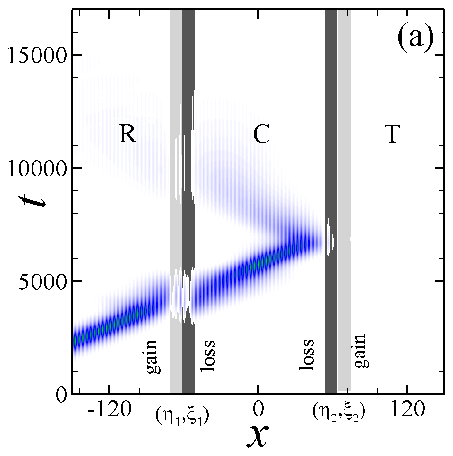,height=5cm}\epsfig{file=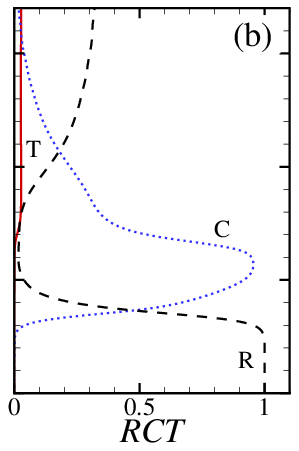,height=5.1cm}
\caption{(Color online) The same as in Fig.~\ref{diode_2} with two different  PT defects with opposite facing loss sides and for parameter values: $x_{1,2}=\pm 20\pi$, $\eta_1=-5.95$, $\xi_{1}/|\eta|=-0.01$ and $\eta_2=-10$, $\xi_{2}/|\eta|=0.01$. Parameters of the initial GS are: $E_s=-0.125$, $v=0.05$.
}	
\label{diode_2_loss}
\end{figure}

Similarly in Fig.~\ref{diode_2_loss} we have considered the case of two PT defects arranged with opposite facing loss sides so that a GS entering via resonant transmission into the intra defect region it becomes  completely depleted by the multiple reflections.
One can also consider an arrangement with the facing sides of the defects having opposite signs so to allow the storage of solitons by compensating the loss  in the reflection at  one side with the gain in the reflection at the other side (not shown here for brevity). PT defect devices based on GS  will be discussed in more details elsewhere.

\section{Conclusions}

In this paper we have  investigated the scattering properties of gap solitons of the periodic nonlinear Schr\"odinger equation (NLSE) in the presence  of localized PT-symmetric defects. The periodic potential responsible for the band-gap structure and for existence of GS has been  taken of trigonometric form while the localized PT-symmetric defect was taken with the real part of Gaussian the and the imaginary part as a product of a Gaussian and a linear ramp potential (antisymmetric in space). We have shown, by means of direct numerical simulations, that by properly designing the amplitudes of real and imaginary parts of the PT defect it is possible to achieve a resonant transmission of the gap soliton through the defect. 
We showed that this phenomenon occurs for potential parameters that support localized modes inside the PT defect potential  with the same energy and  norm of the incoming soliton.  The direct numerical results were found to be  in good agreement with the predictions for the resonant transmission made in terms of stationary defect mode analysis, this extending  previous results for conservative defects~\cite{BS2011} to the case of PT-symmetric  defects. When the imaginary amplitude of the PT defect is increased we found that significant  changes in the scattering properties appear.
In particular, we showed the possibility of  transmitted and reflected GS which gets damped or amplified during the scattering process depending on the side of the defect (loss or gain)  with which the GS interacts more. We investigated this both by means of the mean imaginary part of defect potential seen by the GS and by trajectories followed by the center of the density distribution.  Scattering properties of  gap solitons belonging to different band-gaps and having   different effective masses were also investigated.
We showed that GS with effective masses of opposite sign  behave  similarly in PT defect potentials of opposite sign especially for parameters values close to high resonances. Finally, we discussed the scattering of a GS by a PT defect which leads to an  unidirectional transmission or blockage ({\it diode effect}), and the amplification/depletion of a  GS trapped between a pair of consecutive PT defects.

Finally, in closing this paper we remark that since  PT-symmetric potentials can  be  easily implemented in nonlinear optical systems, we expect the above results to be of experimental interest for systems such as arrays of nonlinear  optical waveguides and photonic crystals.

\section*{Acknowledgements} 
 
M.S. acknowledges support from the
Ministero dell' Istruzione, dell' Universit\'a e della Ricerca
(MIUR) through a {\it Programma di Ricerca Scientifica di
Rilevante Interesse Nazionale} (PRIN)-2010 initiative.
F.A. acknowledges partial support from FAPESP(Brasil).

\end{document}